\documentclass{PoS}

\usepackage{amsmath, amssymb, amsfonts, bbold, subfigure}

\newcommand{\ddd}{\displaystyle}
\newcommand{\bra}[1]{\langle #1|}
\newcommand{\ket}[1]{|#1\rangle}
\newcommand{\vac}{|0\rangle}

\newcommand{\comm}[2]{\left[#1, #2\right]}

\title{
\vspace{-1cm}
\begin{minipage}{\textwidth}
\begin{flushright}
\normalsize PoS(LAT2010)190\\
\normalsize DESY 10-171\\
\normalsize ITEP-LAT/2010-13\\
\end{flushright}
\end{minipage}\\[15pt]
The Chiral Magnetic Effect and chiral symmetry breaking in SU(3) quenched
lattice gauge theory
\thanks{The work was supported by the grant for Leading Scientific Schools NSh-6260.2010.2 and RFBR
 08-02-00661-a, Federal Special-Purpose Programme `Cadres' of the
 Russian Ministry of Science and Education.}
}

\ShortTitle{The Chiral Magnetic Effect and symmetry breaking in SU(3) quenched
theory.}

\author{V.V. Braguta\\
        IHEP, Protvino, Moscow oblast, 142284 Russia\\
        E-mail: \email{braguta@mail.ru}}

\author{P.V. Buividovich\\
        ITEP, B. Cheremushkinskaya str. 25, Moscow, 117218 Russia\\
        JINR, Joliot-Curie str. 6, Dubna, Moscow region, 141980 Russia\\
        E-mail: \email{buividovich@itep.ru}}

\author{\speaker{T. Kalaydzhyan}\\
        DESY Hamburg, Theory Group, Notkestrasse 85, D-22607 Hamburg, Germany\\
        ITEP, B. Cheremushkinskaya str. 25, Moscow, 117218 Russia\\
        E-mail: \email{tigran.kalaydzhyan@desy.de}}

\author{S.V. Kuznetsov\\
        ITEP, B. Cheremushkinskaya str. 25, Moscow, 117218 Russia\\
        E-mail: \email{kuznetsov@itep.ru}}

\author{M.I. Polikarpov\\
        ITEP, B. Cheremushkinskaya str. 25, Moscow, 117218 Russia\\
        E-mail: \email{polykarp@itep.ru}}

\abstract{We study some properties of the non-Abelian vacuum induced by strong
external magnetic field. We perform calculations in the quenched SU(3) lattice
gauge theory with tadpole-improved L\"uscher-Weisz action and chirally
invariant lattice Dirac operator. The following results are obtained: The
chiral symmetry breaking is enhanced by the magnetic field. The chiral condensate depends on the strength of the applied field as a power function with exponent $\nu = 1.6\pm0.2$. There is a
paramagnetic polarization of the vacuum. The corresponding susceptibility and
other magnetic properties are calculated and compared with theoretical
estimations. There are non-zero local fluctuations of the chirality and
electromagnetic current, which grow with the magnetic field strength. These
fluctuations can be a manifestation of the Chiral Magnetic Effect (CME).}

\FullConference{The XXVIII International Symposium on Lattice Field Theory, Lattice2010\\
        June 14-19, 2010\\
        Villasimius, Italy}

\begin{document}

\section{Introduction}

The modern experiments provide a possibility to discover new physical effects
caused by presence of the strong (hadronic scale) magnetic field. At the
Relativistic Heavy Ion Collider (RHIC) at the first moments ($\tau \sim 1$
fm/c) of noncentral collision the very strong ($B \sim 10^{15}$T, $\sqrt{eB}
\sim$ 300 MeV) magnetic fields appear\cite{Kharzeev, Selyuzhenkov}. Such strong
magnetic fields can be also created in ALICE experiment at LHC, at the Facility
for Antiproton and Ion Research (FAIR) at GSI and in the experiment NICA in
Dubna. The additional motivation for the study of the effects induced by the
strong magnetic field could also come from the physics of the early Universe,
where the strong fields ($B \sim 10^{16}$T, $\sqrt{eB} \sim$ 1 GeV) could have
been produced after the electroweak phase transition\cite{Vachaspati}. Due to
the nonperturbative nature of the effects we perform the calculations in the
lattice gauge theory. We use quenched approximation and show that for some
problems it provides rather reasonable values of the physical quantities.

This work has been done analogously to the previous SU(2) lattice
studies\cite{Buividovich_condensate,Buividovich_magnetization,Buividovich_dipole,Buividovich_CME}.
The list of considered effects induced by the magnetic field is the following.

The strong magnetic field can enhance the chiral symmetry breaking. There are
various models (see Sec.\ref{condensatesection}) which predict the growing of
the chiral condensate.

The second effect is the chiral magnetization of the QCD vacuum. This effect
has a paramagnetic nature. The vacuum magnetization is related to the nucleon magnetic moments\cite{Ioffe} and other nonperturbative effects of
hadrons\cite{phenomenology}. We calculate the magnetic susceptibility and other
quantities in Sec.\ref{magnetizationsection}.

The quarks develop an electric dipole moment along the field due to the local
fluctuations of the topological charge\cite{Buividovich_dipole}. We study this
effect in Sec.\ref{dipolesection}.

Finally, the fluctuations of the topological charge can be a source of the
asymmetry between numbers of quarks with different chiralities created in
heavy-ion collisions. The so called ``event-by-event $\mathrm{P}$- and
$\mathrm{CP}$-violation''\cite{Kharzeev} can be explained by this asymmetry and
observed at RHIC. So, our aim is also to see any evidences of this effect in
SU(3) lattice simulations, nevertheless they are similar to SU(2) lattice
results\cite{Buividovich_CME}.

\section{Technical details}

We use the quenched $SU(3)$ lattice gauge theory with tadpole-improved
L\"uscher-Weisz action \cite{Luscher}. To generate the statistically
independent gauge field configurations we use the Cabibbo-Marinari heat bath
algorithm. The lattice size is $14^4$, and lattice spacing $a=0.105fm$. All
observables we discuss later have a similar structure: $\langle\bar\Psi
\mathcal{O} \Psi\rangle$ for VEV of a single quantity or $\langle\bar\Psi
\mathcal{O}_1 \Psi \,\,\, \bar\Psi \mathcal{O}_2 \Psi\rangle$ for dispersions
or correlators. Here $\mathcal{O}$, $\mathcal{O}_1$, $\mathcal{O}_2$ are some
operators in spinor and color space. These expectation values can be expressed
through the sum over $M$ low-lying\footnote{We believe that the IR quantities
are insensitive to the UV cutoff realized by selecting some finite number of
the eigenmodes\cite{Hasenfratz}} but non-zero eigenvalues $i\lambda_k$ of the
chirally invariant Dirac operator $D$ (Neuberger's overlap Dirac
operator\cite{Neuberger}):
\begin{align}
\label{single} \langle\bar\Psi \mathcal{O} \Psi\rangle =
\sum\limits_{|k|<M}\frac{\psi_k^{\dag}\mathcal{O}\psi_k}{i\lambda_k + m}
\end{align}
and
\begin{align}
\label{double} \langle\bar\Psi \mathcal{O}_1 \Psi \,\,\, \bar\Psi \mathcal{O}_2
\Psi\rangle = \sum\limits_{k,
p}\frac{\bra{k}\mathcal{O}_1\ket{k}\bra{p}\mathcal{O}_2\ket{p}-\bra{p}\mathcal{O}_1\ket{k}\bra{k}\mathcal{O}_2\ket{p}}{(i\lambda_k+m)(i\lambda_p+m)},
\end{align}
where all spinor and color indices are contracted and we omit them for
simplicity. The $\lambda_k$ are defined by the equation
\begin{align}
D \psi_k = i \lambda_k \psi_k,
\end{align}
where $\psi_k$ are the corresponding eigenfunctions and the uniform magnetic
field $F_{12}=B_3\equiv B$ is introduced as described
in\cite{Buividovich_condensate}. To perform calculations in the chiral limit
one calculates the expression (\ref{single}) or (\ref{double}) for some
non-zero $m$ and averages it over all configurations of the gauge fields. Then
one repeats the procedure for other quark masses $m$ and extrapolates the VEV
to $m\rightarrow 0$ limit.

\section{Chiral condensate}
\label{condensatesection}

In this section we present our results for the chiral condensate
\begin{align}
\label{condensate}
 \Sigma \equiv -\bra{0} \bar \Psi \Psi \vac,
\end{align}
as a function of the magnetic field $B$. The general tendency for $\Sigma$ to
grow with $B$ was already obtained in various models: in the chiral
perturbation theory \cite{Schramm, Shushpanov} ($\Sigma\propto B$ for weak
fields, $\Sigma\propto B^{3/2}$ for strong fields), in the Nambu-Jona-Lasinio
model \cite{Klevansky} ($\Sigma\propto B^2$), in a confining deformation of the
holographic Karch-Katz model \cite{Zayakin} ($\Sigma\propto B^2$), in D3/D7
holographic system \cite{Kirsch} ($\Sigma\propto B^{3/2}$ for low temperatures,
$\Sigma\propto B$ for high temperatures) and in SU(2) lattice calculations
\cite{Buividovich_condensate}($\Sigma\propto B$). Here our aim is to see how
the chiral condensate behaves in the $SU(3)$ quenched gluodynamics.

\begin{figure}[t]
     \centering
     \subfigure[\label{sigma}]
     {\includegraphics[angle=-90, width=7.5cm]{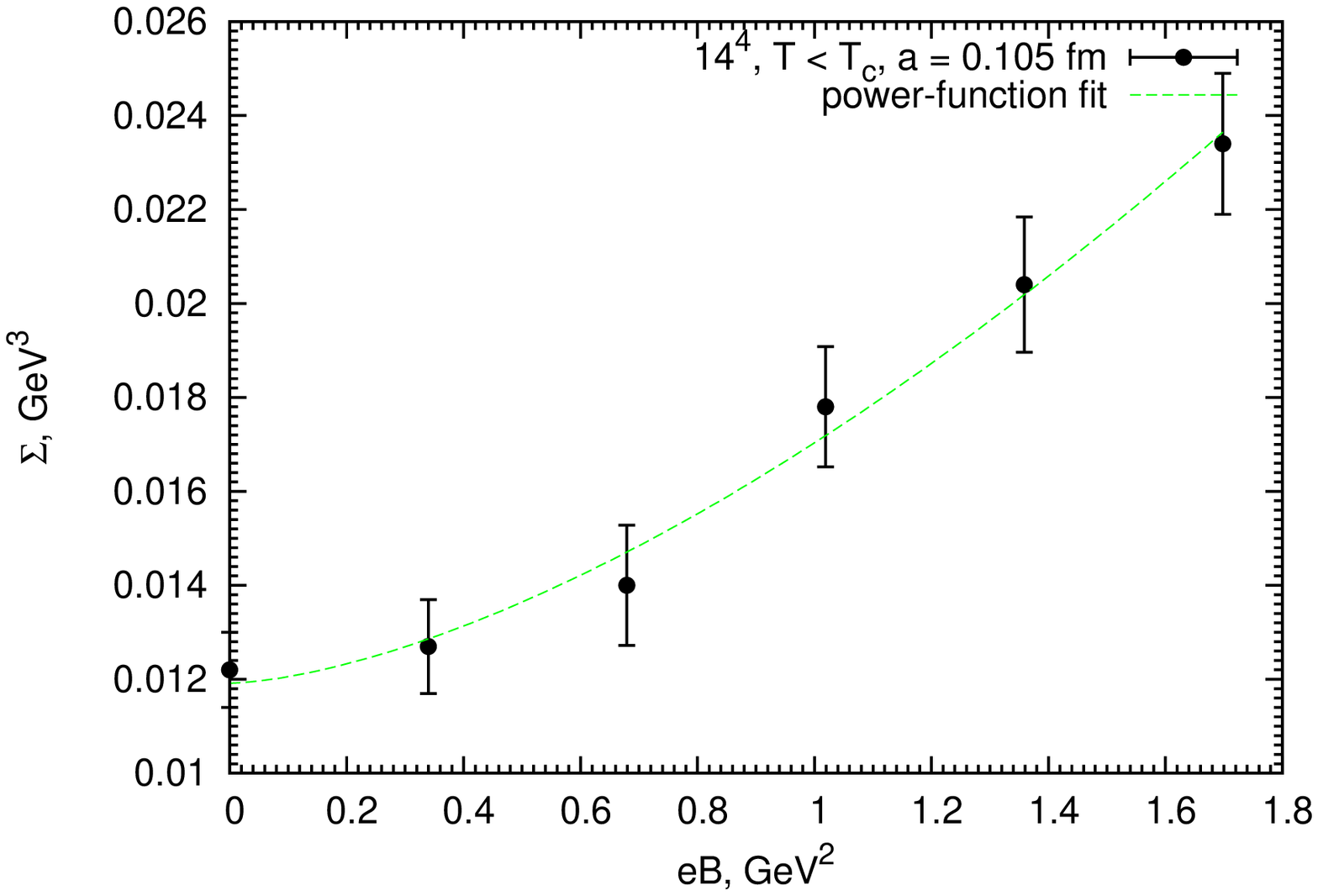}}\hspace{0cm}
     \centering
     \subfigure[\label{sigmaMeV}]
     {\includegraphics[angle=-90, width=7.5cm]{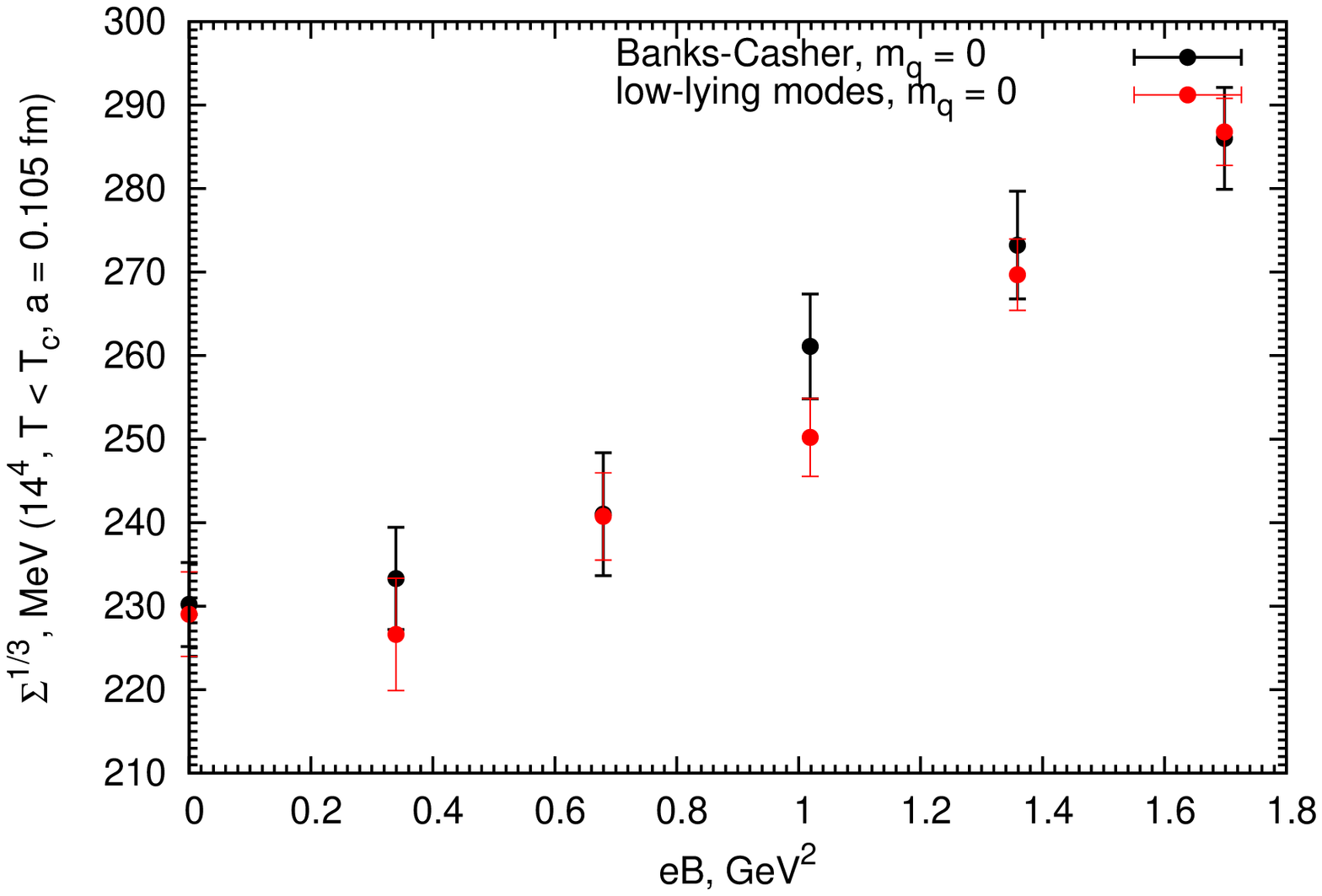}}\hspace{0cm}
\caption{\label{condensateplots} Chiral condensate}
\end{figure}

We use the Banks-Casher formula \cite{Banks_Casher}, which relates the
condensate (\ref{condensate}) with the density $\rho(\lambda)$ of near-zero
eigenvalues of the Dirac operator:
\begin{align}
\label{BCformula} \Sigma = \lim\limits_{\lambda \rightarrow
0}\frac{\pi\rho(\lambda)}{V},
\end{align}
where $V$ is the four-volume of the Euclidean space-time. The result is shown
in Fig.\ref{sigma}.

We perform the fit of the results by the following function:
\begin{align}
 \Sigma^{fit}(B) = \Sigma_0\left[ 1 + \left(\frac{eB}{\Lambda_B^2}\right)^{\nu}\right],
\end{align}
where $\Sigma_0 \equiv \Sigma(0)$. The obtained fitting parameters are
\begin{align}
 \Sigma_0 = \left[(228 \pm 3) MeV\right]^3, \qquad \Lambda_B = \left( 1.31 \pm 0.04 \right) GeV, \qquad \nu = 1.57 \pm 0.23 \,.
\end{align}



It is interesting to compare quantitatively the condensate obtained by the
Banks-Casher formula and that one calculated by the expression (\ref{single})
with $\mathcal{O} = \mathbb{1}$. The result is shown in Fig.\ref{sigmaMeV}. The
value of the condensate in absence of the magnetic field equals $\Sigma(0) =
\left[\left( 230 \pm 5 \right)MeV\right]^3$ which is not so far away from the
value, which can be estimated by the Gell-Mann-Oakes-Renner
formula\cite{Colangelo}:
\begin{align}
\Sigma(0) = \frac{F_{\pi}^2 m_{\pi}^2}{2(m_u + m_d)}\simeq \left[\left(240 \pm
10\right)MeV\right]^3.
\end{align}

\section{Chiral magnetization and susceptibility}
\label{magnetizationsection}

In this section we calculate the quantity
\begin{align}
\label{sigma_def}
 \langle \bar \Psi \sigma_{\alpha\beta} \Psi \rangle = \chi(F) \langle\bar\Psi\Psi\rangle q F_{\alpha\beta},
\end{align}
where $\sigma_{\alpha\beta}\equiv\ddd\frac{1}{2 i}
\comm{\gamma_{\alpha}}{\gamma_{\beta}}$ and $\chi(F)$ is some coefficient of
proportionality (susceptibility), which depends on the field strength.

This quantity was introduced in\cite{Ioffe} and can be used to estimate the
spin polarization of the quarks in external magnetic field. The magnetization
can be described by the dimensionless quantity $\label{mu_def} \mu  = \chi
\cdot qB$, so that
\begin{align}
\label{mu_in_eq} \langle\bar\Psi\sigma_{12}\Psi\rangle = \mu
\langle\bar\Psi\Psi\rangle \,.
\end{align}

\begin{figure}[t]
     \centering
     \subfigure[\label{s}]
     {\includegraphics[angle=-90, width=7.5cm]{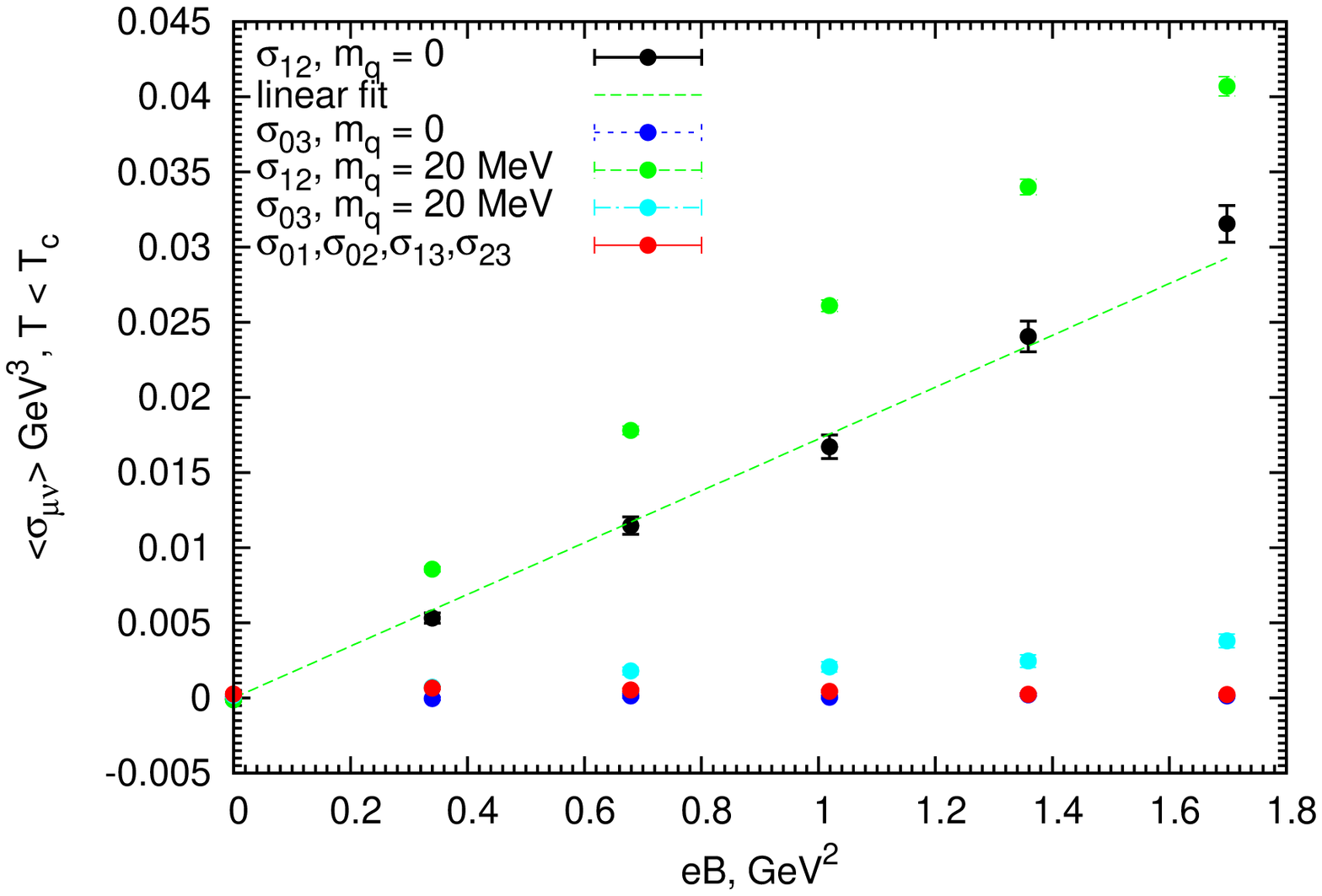}}\hspace{0cm}
     \centering
     \subfigure[\label{s2IR}]
     {\includegraphics[angle=-90, width=7.5cm]{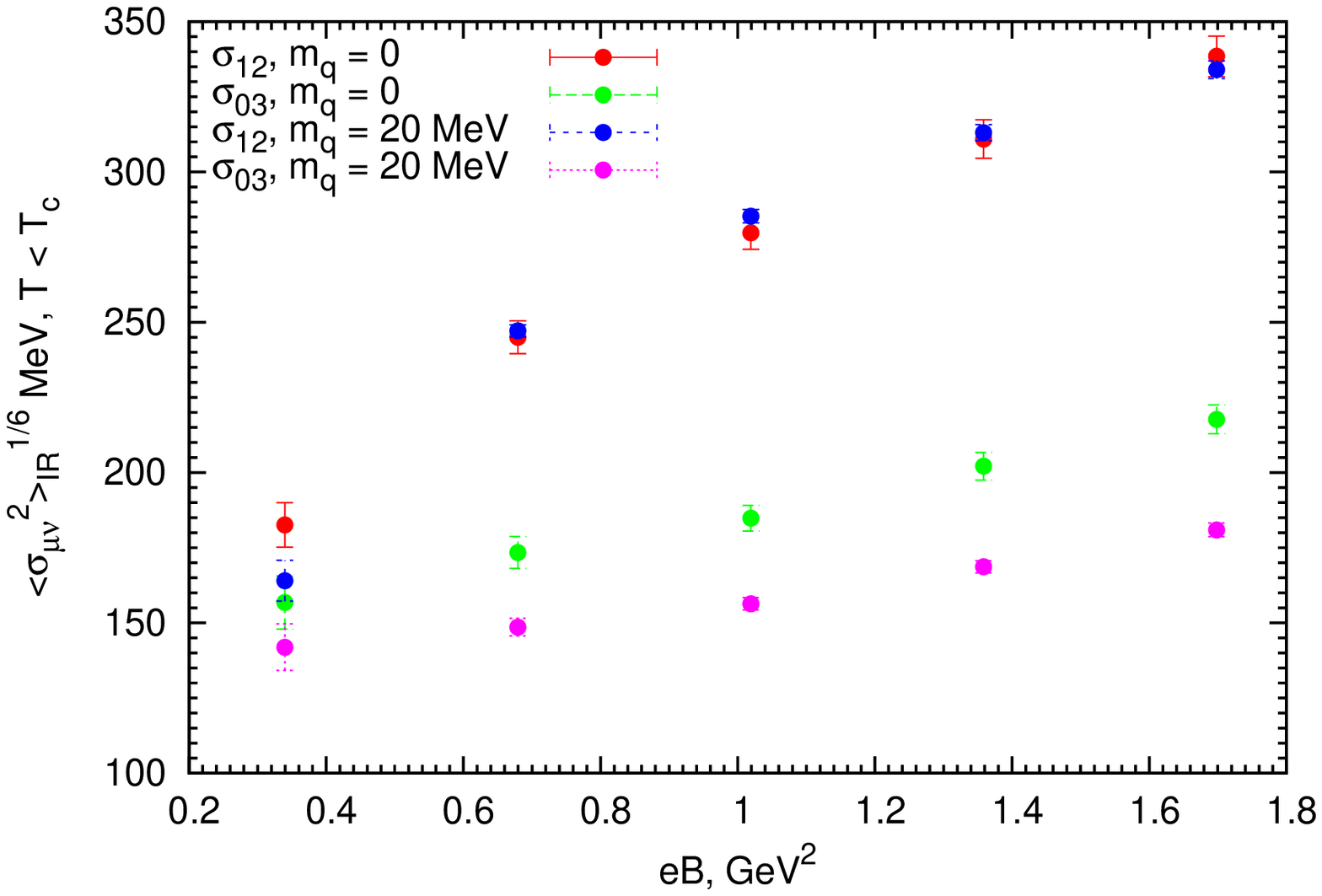}}\hspace{0cm}
\caption{\label{magnetplots} Expectation values of $\bar \Psi
\sigma_{\alpha\beta} \Psi$ and their square}
\end{figure}

The expectation value (\ref{sigma_def}) can be calculated on the lattice by
(\ref{single}) with $\mathcal{O} = \sigma_{\alpha\beta}$. The result is shown
in Fig.\ref{s} (here for comparison we also plot series for some finite quark
mass). We can see, that the 12-component grows linearly with the field, which
agrees with\cite{Ioffe}. This allows us to find the chiral susceptibility
$\chi(0)\equiv\chi_0$. After making a linear approximation
$\langle\bar\Psi\sigma_{12}\Psi\rangle = \Omega^{fit} eB$, where\footnote{in
our simulation we calculate the magnetization of the d-quark condensate, thus
$q=\left|-\frac{e}{3}\right|$}
\begin{align}
\Omega^{fit} \equiv -\frac{1}{3}\chi_0^{fit}\Sigma_0 \,,
\end{align}
we obtain $\Omega^{fit} = (172.3 \pm 0.5) MeV$ and 
\begin{align}
\chi_0^{fit}= -4.24 \pm 0.18\, GeV^{-2}\,.
\end{align}

This value fits well into the range of present theoretical estimations: the modern QCD sum rule calculations \cite{Ball} ($\chi_0^{th}= -3.15 \pm 0.3\, GeV^{-2}$) and \cite{Rohrwild}($-2.85 \pm 0.5\, GeV^{-2}$), the earlier ones \cite{Belyaev} ($-5.7\, GeV^{-2}$) and \cite{Balitsky} ($-4.4 \pm 0.4\, GeV^{-2}$), in the instanton vacuum model \cite{instanton} ($-4.32\, GeV^{-2}$), from the analysis of the Dirac zero-mode in an instanton background \cite{Ioffe:2009} ($-3.52\, GeV^{-2}$), in dubbed quark-meson model \cite{Frasca} ($-4.3\, GeV^{-2}$) and in the Nambu-Jona-Lasinio model \cite{Frasca} ($-5.25\, GeV^{-2}$). We also have to mention a well known analytic result obtained by OPE combined with the idea of pion dominance \cite{Vainshtein} and two  holographic ones \cite{SusGorsky, SusSon}, but for us it seems difficult to compare these results with ours because of an ambiguity in the determination of the pion decay constant $F_\pi$ in the quenched approximation.

Another interesting phenomenological quantity is the product of the chiral
susceptibility $\chi$ and the condensate $\langle\bar\Psi
\Psi\rangle$\cite{phenomenology}. In our calculations it is equal to
\begin{align}
 -\chi_0^{fit}\langle\bar\Psi\Psi\rangle \simeq 52\, MeV,
\end{align}
while from the QCD sum rules one can estimate this quantity as approximately 50
MeV \cite{Ball, Belyaev, Balitsky}, which is also close to our value.

\section{Electric dipole moment}
\label{dipolesection}

Another interesting effect due to the magnetic field is a quark local
electric dipole moment along the field\cite{Buividovich_dipole}. This quantity
corresponds to the $i0$-components of the (\ref{sigma_def}):
\begin{align}
 d_i(x) \equiv \bar \Psi(x) \sigma_{i0} \Psi(x), \qquad\qquad i = \overline{1,3}
\end{align}
In the real CP-invariant vacuum the VEV of this quantity should be zero:
$\langle d_i(x) \rangle = 0$, that we actually see in our results
(Fig.\ref{s}). At the same time the fluctuations of $d_i(x)$ can be
sufficiently strong. We measure VEV's (\ref{double}) with $\mathcal{O}_1 =
\mathcal{O}_2 = \sigma_{\alpha\beta}$. In the case of $i0$-components it
corresponds to dispersions of $\vec{d}$. The result is shown in Fig.\ref{s2IR},
we see that the longitudinal fluctuations of the local dipole moment grow with
the field strength, while transverse fluctuations are absent. Here and after we
use the ``IR'' subscript to emphasize, that we subtract from the quantity its
value at $B=0$:
\begin{align}
\langle Y \rangle_{IR}(B) = \frac{1}{V}\int\limits_V d^4 x\langle Y(x)
\rangle_B - \frac{1}{V}\int\limits_V d^4 x\langle Y(x) \rangle_{B=0}
\end{align}


\section{Some evidences of the Chiral Magnetic Effect}

\begin{figure}[t]
     \centering
     \subfigure[\label{r52IR}]
     {\includegraphics[angle=-90, width=7.5cm]{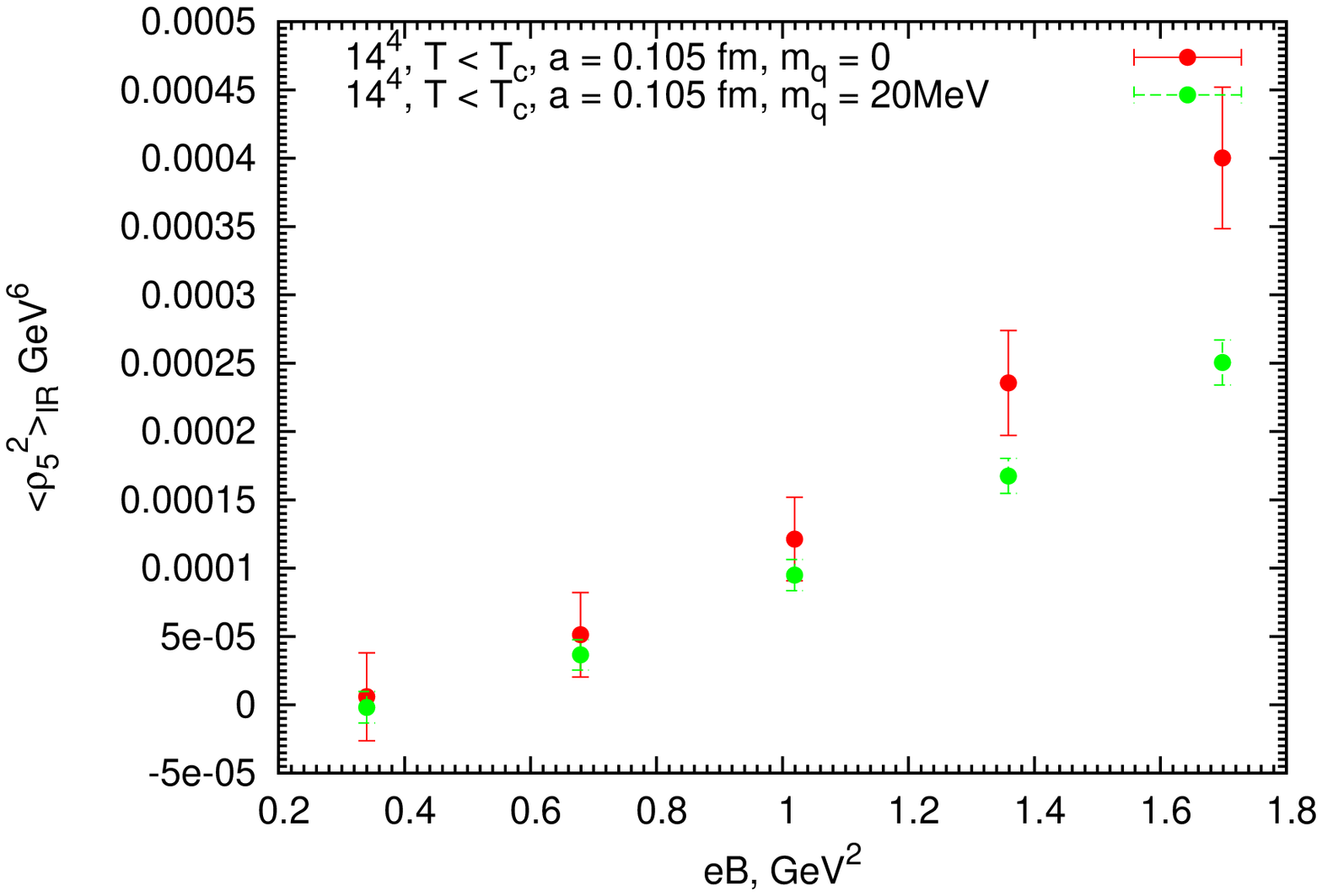}}\hspace{0cm}
     \centering
     \subfigure[\label{j2IR}]
     {\includegraphics[angle=-90, width=7.5cm]{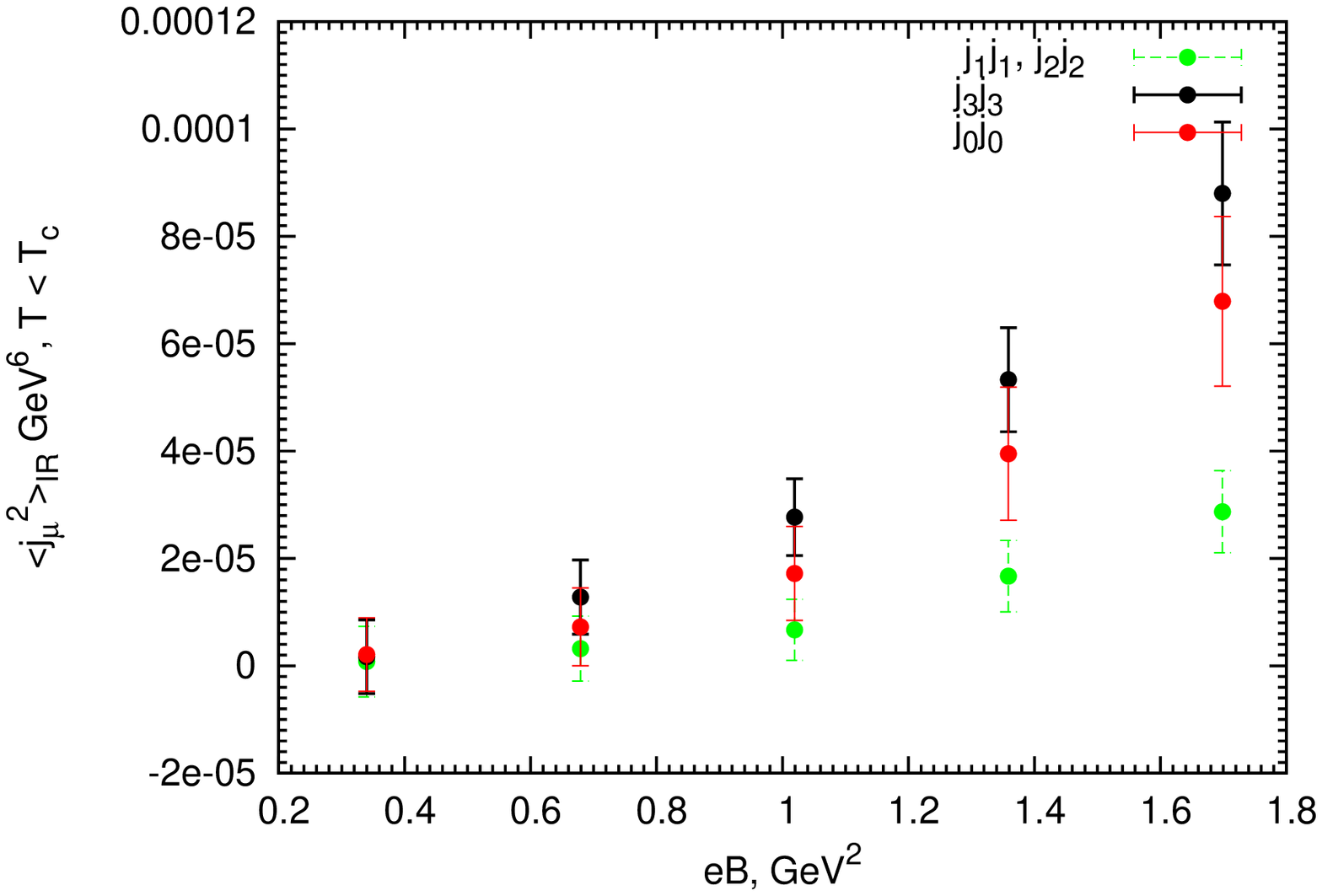}}\hspace{0cm}
\caption{\label{CMEplots} Fluctuations of the chirality and electromagnetic
current/charge}
\end{figure}

The nontrivial topological structure of QCD is due to some nontrivial effects
in the presence of the strong magnetic field. One example of a such effect is
the Chiral Magnetic Effect (CME), which generates an electric current along the
field in the presence of the nontrivial gluonic background\cite{Kharzeev, CME}.
This effect was probably been observed by the STAR collaboration at
RHIC\cite{Voloshin, Caines} in heavy-ion collisions. A lattice evidence of the
effect can be found in\cite{Buividovich_CME, conductivity,Buividovich_recent}.
Here we implement the same procedure for the $SU(3)$ case and study the local
chirality
\begin{align}
 \rho_5(x) = \bar\Psi(x) \gamma_5 \Psi(x) \equiv \rho_L(x) - \rho_R(x)
\end{align}
and the electromagnetic current
\begin{align}
 j_{\mu}(x) = \bar\Psi(x) \gamma_{\mu} \Psi(x).
\end{align}
The expectation value of the first quantity can be computed by (\ref{single})
with $\mathcal{O} = \gamma_5$ and with $\mathcal{O} = \gamma_{\mu}$ for the
second quantity. The both VEV's are zero, as expected, but the corresponding
fluctuations obtained from (\ref{double}) are finite and grow with the field
strength (see Fig.\ref{CMEplots}).

\end{document}